\documentstyle[12pt,epsfig]{article}
\begin{document}

\begin{titlepage}
\rightline{September 2012}
\rightline{Revised: June 2013}
\vskip 2cm
\centerline{\large \bf
Hidden sector dark matter explains the}
\vskip 0.5cm
\centerline{\large \bf 
DAMA, CoGeNT, CRESST-II and CDMS/Si experiments}

\vskip 2.2cm
\centerline{R. Foot\footnote{
E-mail address: rfoot@unimelb.edu.au}}

\vskip 0.7cm
\centerline{\it ARC Centre of Excellence for Particle Physics at the Terascale,}
\centerline{\it School of Physics, University of Melbourne,}
\centerline{\it Victoria 3010 Australia}
\vskip 2cm
\noindent
We examine data from the DAMA, CoGeNT, CRESST-II and CDMS/Si direct detection experiments  
in the context of 
multi-component hidden sector dark matter. The models considered
feature a hidden sector with two or more stable particles charged under an unbroken 
$U(1)'$ gauge interaction. 
The new gauge field can interact with
the standard $U(1)_Y$ via renormalizable kinetic mixing, leading to Rutherford-type elastic
scattering of the dark matter particles off ordinary nuclei.
We consider the simplest generic model of this type, with a hidden sector composed of two stable 
particles, $F_1$ and $F_2$. 
We find that this simple model
can simultaneously explain the DAMA, CoGeNT, CRESST-II and CDMS/Si data. 
This explanation has some  
tension with the most recent results from the XENON100 experiment. 

\end{titlepage}

\section{Introduction}

The existence of non-baryonic dark matter, is clearly indicated by observations on
small and large astronomical scales. In particular, the observed flat rotation curves
in spiral galaxies, CMB anisotropies, and matter power spectrum suggest the existence of 
a cold (or warm) dark matter
component with energy density around five times larger than ordinary baryons\cite{reviewdm}.    
These inferences aside,
the precise nature of dark matter is not yet known. If dark matter
interacts with ordinary matter then it might be directly detectable with Earth based experiments, some of which
have yielded some impressive results. Most notable, are
the DAMA/NaI \cite{dama1} and DAMA/LIBRA \cite{dama2} experiments. 
These experiments have observed an annual modulation 
in the `single hit' event rate, 
over more than 12 annual cycles, with phase consistent with expectations 
from dark matter interactions\cite{dm}.  
Over the last few years, the CoGeNT experiment\cite{cogent,cogent2}, 
using a p-type point-contact Germanium detector with very low energy threshold,
has observed a rising event rate at low energies. These events could not be explained
by known backgrounds and provide evidence supporting the DAMA signal.
Most recently, the CRESST-II\cite{cresst-II} and CDMS/Si\cite{cdms} experiments have obtained results
that are  also compatible with a dark matter signal rising at low energies.

A satisfactory explanation of these experiments requires a specific model for dark matter.
One promising approach appears to be in the framework of multi-component hidden
sector models.  In this scenario
dark matter arises from a multi-component hidden sector which contains an unbroken
$U(1)'$ gauge interaction.
The new gauge field is presumed to interact with
the standard $U(1)_Y$ via renormalizable kinetic mixing, leading to Rutherford-type elastic
scattering of the dark matter particles off ordinary nuclei.
The specific case where the hidden sector
is isomorphic to the ordinary sector, mirror dark matter\cite{flv,review}, has also been discussed at length,
and found to be compatible with the DAMA, CoGeNT, CRESST-II and CDMS/Si 
results\cite{foot2012, foot69, footold,foot2008,foot1}. 
More generally, though, these works indicate that there
is a larger class of such hidden sector models which are capable of explaining
the experiments. 
In this article we examine in detail the simplest such multi-component 
hidden sector model. It serves as a useful prototype of 
generic dark matter models of this kind.

The outline of this article is as follows. Section 2 reviews some general aspects of the 
chosen hidden sector framework. Section 3 examines the specific case of the model
with two stable hidden sector particles, $F_1$ and $F_2$. 
Section 4 confronts this two component hidden sector model with the latest data from DAMA, CoGeNT, 
CRESST-II and CDMS/Si experiments. Section 5, follows up on an example near the combined best
fit identified in section 4.  The compatibility of this example
with XENON100 data is looked at in section 6. Finally, in section 7 we conclude.

\newpage

\section{Hidden sector dark matter}

The models considered relegate dark matter particles to a generic hidden sector
with an unbroken $U(1)'$ gauge interaction.  The special case where the hidden sector
is isomorphic to the standard model sector - mirror dark matter - has been discussed
in the context of direct detection experiments in ref.\cite{foot2012,foot69,footold,foot1}.
Here we examine the more generic hidden sector case. This  includes models with broken
mirror symmetry\cite{chinese} and potentially many other hidden sector models.
Somewhat related models have also been studied in the recent literature e.g.\cite{feng}.
More generally, hidden sector models with light stable particles can also be motivated
from the similar inferred energy density of the ordinary and dark matter in the Universe, see e.g.\cite{kali}.
Although cosmological and astrophysical discussions are beyond the scope of this paper, we 
refer the reader 
to recent work\cite{50} which indicates that
dissipative hidden sector dark matter 
can be rich enough to explain the puzzling regularities observed in spiral galaxies (cored
dark matter density profile etc)\cite{50}.
Also, such models might  potentially be consistent with observations of  
elliptical galaxies\cite{50}, the bullet cluster\cite{sil,50} and other astrophysical
observations, e.g.\cite{footlet}.

We consider spin-independent elastic scattering of dark matter particles on ordinary nuclei.
The dark matter particles are assumed to arise from a hidden sector which contains an unbroken $U(1)'$ gauge
interaction and possibly other gauge interactions.  This means that the interactions of the theory are
described by the Lagrangian:
\begin{eqnarray}
{\cal L} = {\cal L}_{SM} (e, \nu, u, d, B_\mu, ...) + {\cal L}_{dark} (F_1, F_2, .., F_N, B'_\mu, ...) 
+ {\cal L}_{mix}
\end{eqnarray}
where ${\cal L}_{mix}$ contains possible interactions connecting ordinary and hidden sector
particles.
An example of ${\cal L}_{dark}$, with $N$ Dirac fermions $F_j, \ j=1,...,N$, is
\begin{eqnarray}
{\cal L}_{dark} =  - {1 \over 4} F'_{\mu \nu} F^{' \mu \nu} + \bar F_j \left( iD_{\mu}\gamma^\mu - m_j
\right) F_j 
\label{mary}
\end{eqnarray}
where $D_\mu = \partial_\mu + ig'B'_\mu Q'$ is the covariant derivative and $j$ is summed over $1,...,N$.
In this example the $N$ particles $F_j$ are each absolutely stable due to the presence 
of $U(1)_1\otimes U(1)_2 \otimes ....\otimes U(1)_N$ global symmetries. Importantly these 
symmetries are not imposed; they are accidental, like the $U(1)_B$ and $U(1)_L$ global
symmetries of the standard model. 
Evidently, having dark matter arise from a hidden sector is theoretically attractive because it is a natural way to introduce
dark matter particles which are dark, massive, stable and importantly do not modify standard model physics. 

Dark matter elastic scattering depends on the cross-section, $d\sigma/dE_R$, and on the halo
distribution function, $f_{F_j}$, of the dark matter species (taking the general case
of $N$ stable species).  Both of these quantities depend crucially
on the particle physics. Considering first the cross-section, this can have different
recoil energy dependence depending
on the type of interaction.  The most common form discussed in the literature is contact interactions which give
an energy independent cross-section (excepting here the
energy dependence due to the form factor). However, we consider Rutherford scattering which
features non-trivial energy dependence, $d\sigma/dE_R \propto 1/E_R^2$\cite{foot69}.
From the point of view of hidden sector dark matter, where the hidden sector contains
an unbroken $U(1)'$ gauge interaction, the latter appears
to be the most natural form of interaction since it arises from $U(1)'-U(1)_Y$ 
kinetic mixing which is
both gauge invariant and renormalizable.
Rutherford scattering results because the interaction is mediated by a massless particle (the photon).
\footnote{
The case where the mediator is light is also possible, and has been studied recently in ref.\cite{panci}.
As shown there,
in the context of the DAMA, CoGeNT and CRESST-II experiments,
Rutherford scattering arises if the mediator has mass much less than 10 MeV, while point-like
interaction arises if the mediator has mass much greater than 10 MeV.}

The distribution function, $f_{F_j}$, also depends on the particle physics. For dark 
matter composed of only one type of particle with mass, $m$, the distribution is typically assumed
to be Maxwellian, with an effective temperature set by the galactic rotational 
velocity: $T  \approx {1 \over 2} m v_{rot}^2$.
This leads to a distribution: $f = exp(-v^2/v_0^2)$ with $v_0 \approx v_{rot}$.
On the other hand, if dark matter is multi-component and self-interacting both plausible if
dark matter arises from a hidden sector with unbroken $U(1)'$ gauge interaction, then the distribution
function is still Maxwellian, but the velocity dispersion depends on the mass
of the species.
$v_0[F_j] = \sqrt{2T/m_j}$ (recall the index $j$ labels the stable particle species, $j=1,...,N$). 
If the interactions are frequent enough so that the particles
in the halo have a common (local) temperature\footnote{
This assumes that the time scale for self interactions in the Milky Way galaxy 
is much shorter than cosmological time scales.
This assumption is generally valid if the interactions have similar strength
to those of the ordinary charged particles, see e.g.\cite{footlet}.},
$T$,
then it follows that the velocity dispersion
depends on $m_j$ via $v_0[F_j] \propto 1/\sqrt{m_j}$. In fact assuming 
an isothermal pressure supported halo
in hydrostatic equilibrium,
one can show that\cite{footold}  
\begin{eqnarray}
v_0[F_j] = v_{rot} \sqrt{{\bar m \over m_j}}\ 
\label{v0}
\end{eqnarray} 
where $\bar m = \sum n_j m_j/\sum n_j$ is the mean mass of the particles in the halo.
Thus the velocity dispersion can vary greatly depending on the masses within the model.

\section{Two component hidden sector dark matter}

Consider the case where the hidden sector contains two stable $U(1)'$ charged 
dark matter particles $F_1$ and $F_2$
with masses $m_{F_1}$ and $m_{F_2}$ [$F_1$ and $F_2$ can be fermionic as in the example Eq.(\ref{mary}) 
or alternatively bosonic].
If their abundance in the Universe arises from a particle-antiparticle asymmetry
then two stable particles with opposite
in sign,
but not necessarily equal in magnitude, 
$Q'$ charges is actually the minimal case given $Q'$ neutrality of the Universe.
In fact this neutrality is not just global, $U(1)'$ electric fields will ensure local neutrality 
of a plasma containing $F_1$ and $F_2$ particles.
That is,
\begin{eqnarray}
n_{F_1} Q'_{F_1} + n_{F_2}Q'_{F_2} = 0
\label{mary5}
\end{eqnarray}
where $n_{F_j}$ [$Q'_{F_j}$] is the local number density [$U(1)'$ charge] of $F_j$.

We assume that the $U(1)'$ gauge field interacts
with the standard $U(1)_Y$ gauge field via the gauge invariant and renormalizable kinetic mixing 
interaction\cite{he}:
\begin{eqnarray}
{\cal L}_{mix} = \frac{\epsilon'}{2\cos\theta_w} F^{\mu \nu} F'_{\mu \nu}
\label{kine}
\end{eqnarray}
where $F_{\mu \nu}$ is the standard $U(1)_Y$ gauge boson 
field strength tensor, and $F'_{\mu \nu}$ is the field strength tensor for the hidden sector $U(1)'$. 
This interaction enables the hidden sector $U(1)'$ charged particles $F_j$ to couple to
ordinary photons\cite{holdom} with electric charge $g'Q'_{F_j} \epsilon' \equiv \epsilon_{F_j} e$.
[Henceforth we define $\epsilon \equiv \epsilon_{F_2}$].
The cross-section of a $F_2$ particle 
to elastically scatter off an ordinary nucleus, $A$, presumed at rest with  
atomic number $Z$, is given by\cite{foot69}:\footnote{Unless otherwise indicated, 
natural units where $\hbar = c = 1$ are assumed.}
\begin{eqnarray}
{d\sigma \over dE_R} = {\lambda \over E_R^2 v^2}
\label{cs}
\end{eqnarray}
where 
\begin{eqnarray}
\lambda \equiv {2\pi \epsilon^2 Z^2 \alpha^2 \over m_A} F^2_A (qr_A)   \
\end{eqnarray}
and $F_A (qr_A)$ is the form factor which
takes into account the finite size of the nuclei.
Helm proposed a simple analytic expression for 
the form factor, which we adopt in our numerical work\cite{helm,smith}.

Rotation curve data in
spiral galaxies suggests that the  $F_1$ and $F_2$ particles  are (roughly) 
spherically distributed  
in a pressure supported halo [c.f. ref.\cite{sph}].
We assume that the self-interactions are frequent enough as to thermalize the
distributions of $F_1$ and $F_2$ with a common temperature $T$. 
We further assume that 
the binding energy of atomic bound states containing $F_1$ and $F_2$
particles are 
much less than this temperature, so that the $F_1$ and $F_2$ particles can be treated as two components of a plasma
\footnote{The alternative case, where $F_1$ and $F_2$ are tightly bound into atoms has been
discussed recently in ref.\cite{cline}. In the tightly bound limit, the interaction between ordinary matter
and such atomic dark matter becomes point-like, and its implications for direct detection experiments
are quite different from the case we study here.}.
Typically this requires $\alpha'^2 m_{F_1}m_{F_2}/(m_{F_1}+m_{F_2}) \ll {\rm keV}$ where $\alpha' \equiv g'^2 Q'_{F_1} Q'_{F_2}/4\pi$.

As mentioned earlier, the condition of hydrostatic
equilibrium relates the temperature of the particles to
the galactic rotational velocity, $v_{rot}$, resulting in a mass dependent 
velocity dispersion, Eq.(\ref{v0}).
In the two component case the mean mass of the particles in the galactic halo is given by
\begin{eqnarray}
\bar m &\equiv & {n_{F_1} m_{F_1} + n_{F_2} m_{F_2} \over n_{F_1} + n_{F_2}} \nonumber \\
& = & {m_{F_1} - {Q'_{F_1} \over Q'_{F_2}} m_{F_2} \over 1 - {Q'_{F_1} \over Q'_{F_2}}}
\end{eqnarray}
where we have made use of Eq.(\ref{mary5}).
From the point of view of direct detection experiments
the interesting region of parameter space 
is where $m_{F_1} \ll m_{F_2}$ and $|Q'_{F_1}| \ll |Q'_{F_2}|$ so that $\bar m \ll m_{F_2}$. 
It then follows from Eq.(\ref{v0}) that $v_0^2 (F_2) \ll v_{rot}^2$. 
The narrow velocity dispersion (recall $\sigma_v^2 = 3v_0^2/2$) can greatly reduce the rate
of $F_2$ interactions in higher threshold experiments such as XENON100\cite{xenon100} and CDMS/Ge
\cite{cdmsge} whilst still allowing $F_2$ 
to explain the signals in the lower threshold DAMA and CoGeNT experiments.  
The $F_1$ state can be too light to be directly detected in the experiments 
(roughly this means that $m_{F_1} \stackrel{<}{\sim} 5$ GeV),
but influences the way $F_2$ interacts due to its effect on the velocity dispersion of $F_2$.
With these assumptions,
current experiments depend on three parameters, $m_{F_2}, \bar m, \epsilon\sqrt{\xi_{F_2}}$ 
(the parameter, $\xi_{F_2}$ will be defined in a moment).
This two component hidden sector dark matter model
has been briefly discussed previously in ref.\cite{foot1}. Our purpose here is to study
it in more detail taking into account the tentative dark matter signal coming from the CRESST-II and CDMS/Si experiments\cite{cresst-II} as well
as the updated results from CoGeNT\cite{cogent2}.

The differential scattering rate for $F_j$ on a target nuclei, $A$, is given 
by\footnote{The upper limit of integration in Eq.(\ref{55}) 
is taken as infinity since we are dealing with
dark matter particles with potentially significant self-interactions. 
The self-interactions can prevent
particles in the high velocity tail of the Maxwellian distribution from escaping the galaxy. }:
\begin{eqnarray}
{dR \over dE_R} = N_T n_{F_j} 
\int^{\infty}_{|{\textbf{v}}| > v_{min}}
{d\sigma \over dE_R}
{f_{F_j}({\textbf{v}},{\textbf{v}}_E) \over k} |{\textbf{v}}| d^3v 
\label{55}
\end{eqnarray}
where the integration limit, $v_{min}$, is defined by the kinematic relation:
\begin{eqnarray}
v_{min} &=& \sqrt{ {(m_{A} + m_{F_j})^2 E_R \over 2 m_{A} m^2_{F_j} }}\ .
\label{v}
\end{eqnarray}
In Eq.(\ref{55}), $N_T$ is the number of target nuclei per kg of detector and   
$n_{F_j}$ is the number density of the halo dark matter particles $F_j$ at the Earth's
location. This number density can be expressed in terms of the halo
mass fraction of $F_j$, $\xi_{F_j}$, and total mass density, 
$\rho_{dm}$ via $n_{F_j}  = \rho_{dm} \xi_{F_j}/m_{F_j}$ 
(we set $\rho_{dm} = 0.3 \  {\rm GeV/cm}^3$).
Also in Eq.(\ref{55}) ${\bf{v}}$ denotes the velocity of the halo particles relative to the
Earth and ${\bf{v}}_E$ the
velocity of the Earth relative to the galactic halo\footnote{In all numerical work
we include an estimate of the Sun's peculiar velocity so that  
$\langle |{\bf{v}}_E| \rangle = v_{rot} + 12$ km/s.}.
The halo distribution function, in the reference frame of the Earth, is then given by
the Maxwellian distribution:
\begin{eqnarray} 
{f_{F_j} ({\bf{v}},{\bf{v}}_E) \over k} = (\pi v_0^2[F_j])^{-3/2} exp\left({-({\bf{v}}
+ {\bf{v}}_E)^2 \over v_0^2[F_j]}\right)\ . 
\label{pw}
\end{eqnarray}
The integral, Eq.(\ref{55}), can be simplified in terms
of error functions\cite{foot2008,smith} and solved numerically.
Detector resolution effects can be incorporated by convolving the rate with 
a Gaussian, as was done for the mirror dark matter case in\cite{foot2012}. 


\section{Direct detection of hidden sector dark matter}

As discussed earlier, the two component hidden sector model supposes dark matter arises from
a hidden sector with two components $F_1$ and $F_2$. With $m_{F_1} \stackrel{<}{\sim} 5$ GeV,
the existing direct detection experiments depend on three parameters: $m_{F_2}, \ 
\epsilon \sqrt{\xi_{F_2}}, \ \bar m$.
It was shown in ref.\cite{foot1} that with $\bar m = 1$ GeV and $v_{rot} = 240$ km/s (fixed as a specific example) 
this model could provide a reasonable fit to the
DAMA annual modulation signal and measured CoGeNT spectrum. 
Since that time, the measured CoGeNT spectrum has undergone a significant `surface event correction'\cite{cogent2}.
Additionally, CRESST-II and CDMS/Si have announced results suggesting a tentative
dark matter signal.  Our purpose now is to
re-examine the two component hidden sector model in light of these new experimental developments.
We shall also extend our previous analysis by studying
a wider range of $\bar m, \ v_{rot}$ values.
In particular we take $\bar m/{\rm GeV} = 1.0$  and $3.0$ and
two values for the rotational velocity, $v_{rot} = 200$ km/s and
$v_{rot} = 240$ km/s.
Although $v_{rot} = 240$ km/s is representative of recent
measurements of the local rotational velocity of the Milky Way\cite{rot} there are significant
uncertainties in this quantity and it is useful to see how things change when
$v_{rot}$ is varied.

\vskip 0.7cm
\noindent
{\it The CRESST-II experiment}
\vskip 0.4cm

The CRESST-II experiment has collected a $730$ kg-day exposure of a $Ca W O_4$ target\cite{cresst-II}.
The data arises from eight detector modules, with 
recoil energy thresholds (keV) of $10.2, 12.1, 12.3, 12.9, 15.0, 
15.5, 16.2, 19.0$.
The binned CRESST-II data is given in table 1.
For the CRESST-II analysis, we define $\chi^2$: 
\begin{eqnarray}
\chi^2 (m_{F_2}, \epsilon \sqrt{\xi_{F_2}}, \bar m) = 
\sum_{i=1}^{5}  \left[ {R_i + B_i - data_i \over \delta data_i}\right]^2 \ 
\label{chi2bla}
\end{eqnarray}
where $R_i$ is the predicted rate and $B_i$ is the estimated
background in the $i^{th}$ energy bin.
The relevant rates, $R_i$, for the CRESST-II experiment are computed 
from Eq.(\ref{55}).
The resolution is included using 
$\sigma_{res}  = 0.3$ keV\cite{cresst-II}. Also included are the
detection efficiencies, $\epsilon_f$, for the three target elements: 
$\epsilon_f = 0.9$ for $O, W$ and
$\epsilon_f = 1.0$ for $Ca$, which take into account their acceptance region.
No energy scale uncertainty is considered for CRESST-II.

\begin{table}
\centering
\begin{tabular}{c c c}
\hline\hline
Bin / keV & Total events & Estimated background  \\
\hline
10.2 -- 13.0 & 9 & 3.2  \\
13 -- 16 & 15 & 6.1 \\
16 -- 19 & 11 & 7.0  \\
19 -- 25 & 12 & 11.5  \\
25 -- 40 & 20 & 20.1 \\
\hline\hline
\end{tabular}
\caption{CRESST-II data: total number of events and estimated background.}
\end{table}

\vskip 0.7cm
\noindent
{\it The DAMA experiment}
\vskip 0.4cm

The DAMA/NaI and DAMA/LIBRA experiments have accumulated data 
from a large exposure [$1.17$ ton-year] of a $Na I$ target\cite{dama1,dama2}. 
Analysis of this data has yielded an annual modulation with phase consistent with dark matter
expectations at around $8.9\sigma$ C.L.
We analyse this annual modulation signal
in the energy range $2-8$ keVee
using 12 bins of width $0.5$ keVee\cite{dama2}.
The detector resolution, obtained from ref.\cite{damares}, has been included.
The quenching factors
$q_{Na}, \ q_I$, which set the nuclear recoil energy scale,
are uncertain [${\rm keVee} = {\rm keV}_{NR}/q$]. 
Following the mirror dark matter analysis of \cite{foot2012},
we consider a range of possible values for the quenching factors:
\footnote{
In our analysis we neglect the possibility of channeling 
(scatterings with $q \simeq 1$). A theoretical study\cite{newstudy}
and also recent experimental work\cite{collarchan} both suggest
that the channeling fraction is probably small. Nevertheless we should
keep in mind that 
even a channeling fraction as low as a few percent 
can significantly shift the DAMA favored regions of parameter space to lower
$\epsilon \sqrt{\xi_{F_2}}$ values.} 
\begin{eqnarray}
q_{Na} = 0.28 \pm 0.08, \ \ 
q_I = 0.12\pm 0.08 
\ .
\label{moomba}
\end{eqnarray}
Assuming that $q_I, \ q_{Na}$ are energy independent,
we minimize $\chi^2$ varying  $q_{Na}, \ q_I$ over the above range.
Values of $q_I, \ q_{Na}$ outside this range are possible and have been
discussed in the literature. For example,
the higher values $q_{Na} \approx 0.6, \ q_I \approx 0.3$
have been
suggested by Tretyak\cite{tretyak}.  Although 
the possibility of such high quenching factors are not specifically considered in our
numerical work,
we note here that higher values for the quenching factors generally move
the DAMA allowed region to lower values of $\epsilon \sqrt{\xi_{F_2}}, \ m_{F_2}$.

\vskip 0.5cm
\noindent
{\it The CoGeNT experiment}
\vskip 0.4cm

The CoGeNT collaboration has recently presented data corresponding to an exposure of 
$0.33\times 807$ kg-days in a Germanium target\cite{cogent2}. This update of their earlier 
exposure\cite{cogent} includes an important correction for surface events which have not been excluded
by their rise time cut\cite{cogent2}.
The efficiency corrected and surface event corrected CoGeNT data
is obtained from figure 21 of ref.\cite{cogent2}.
This data is analysed using 15 bins of width $0.1$ keVee in the region $0.5-2.0$ keVee 
taking into account the advertised detector resolution.
We allow for a constant background, which we fit to the data in this energy
range.  Uncertainties in energy scale are included by
minimizing the $\chi^2$ for CoGeNT over the variation in quenching factor,
$q_{Ge} = 0.21 \pm 0.04$.

\vskip 0.5cm
\noindent
{\it The CDMS/Si experiment}
\vskip 0.4cm

The CDMS/Si experiment has observed three dark matter
candidate events in an array of silicon detectors\cite{cdms}.
These three events have nominal recoil energies: 8.2 keV, 9.5 keV and 12.3 keV.
A $\chi^2$ analysis cannot be used given the low number of events.
Instead, the likelihood function is constructed using the extended maximum likelihood formalism\cite{barlow}.
This has the form:
\begin{eqnarray}
{\cal L}({\bf p}) = \left[ \Pi_{i} {dn (E_R^i) \over dE_R} \right] exp[-{\cal N({\bf p})}]
\end{eqnarray}
where the unknown parameters are denoted by the vector ${\bf p}$.
Here, $dn (E_R^i)/dE_R$
is the expected event rate evaluated at the recoil energy for the three observed events, $i=1,...,3,$.
The total number of expected events is 
\begin{eqnarray}
{\cal N({\bf p})} = \int {dn \over dE_R} \ dE_R
\end{eqnarray}
where the integration limits cover the acceptance recoil energy region.
The expected event rate, $dn/dE_R$, is the rate $dR/dE_R$, Eq.(\ref{55}),
convolved with a Gaussian to take into account the resolution\footnote{
In the absence of resolution measurements, we take $\sigma_{res} = 0.1$ keV.}
and
multiplied by the detection efficiency, $\epsilon_f (E_R)$ (obtained from
ref.\cite{cdms} for the 140.2 kg-day exposure).

There are some indications\cite{cdms} that the recoil energy calibration is
around 10\% higher than nominally used, with some uncertainty. We have thus scaled up the energies
by a factor, $f=1.1$ and considered an energy calibration uncertainty of $\pm 10\%$, i.e. $f = 1.1 \pm 0.1$.
The profile likelihood function, ${\cal L}_P$,
is then obtained by
maximizing ${\cal L}$ over this
range of $f$ for each  
value of the parameters: $m_{F_2}, \ \epsilon \sqrt{\xi_{F_2}}$. 
The favored region for the parameters $m_{F_2},\ \epsilon \sqrt{\xi_{F_2}}$
can then be obtained from the condition:
\begin{eqnarray}
ln \ {\cal L}_P \ge ln \ {\cal L}_{P max} - \Delta \ ln \ {\cal L}_P
\ .
\end{eqnarray}
We set
$2\Delta \ ln \ {\cal L}_p = 5.99 $ which corresponds to 95\% C.L. for 2 parameters\cite{revppg}.
In this analysis we
neglect the background contribution. This is justified given that the total background is 
estimated to be much less than 1 event for the CDMS exposure (in the
energy region of interest, $E_{threshold} \le E_R \le 20$ keV)\cite{cdms}.

\vskip 0.4cm
\noindent
{\large \it The analysis}
\vskip 0.4cm

We summarize the $\chi^2 (min)$ values for the relevant data sets from 
each experiment for the two chosen $\bar m$ values in table 2 for 
$v_{rot} = 200$ km/s and table 3 for $v_{rot} = 240$ km/s.
The 95\% C.L. favored region of parameter
space is bounded by the contours where
$\chi^2 (m_{F_2}, \epsilon \sqrt{\xi_{F_2}}) = \chi^2_{min} + 5.99$.
In figure 1 we plot the favored region of parameter space for
DAMA, CoGeNT, CRESST-II and CDMS/Si for $v_{rot} = 200$ km/s and with reference
value of $\bar m $: $\bar m = 1.0$ GeV (figure 1a), $\bar m = 3.0$ GeV (figure 1b).
In figure 2 we repeat the exercise, but with $v_{rot} = 240$ km/s.

\begin{table}
\centering
\begin{tabular}{c c c c}
\hline\hline
Experiment & $\chi^2 (min)/d.o.f. $ & $\bar m$/GeV & Best fit parameters  \\
\hline
DAMA (annual mod.) & $5.8/10$ & $\bar m = 1.0$ & $m_{F_2} = 56.0$ GeV, 
$\epsilon \sqrt{\xi_{F_2}} = 7.2 \times 10^{-9}$  \\
CoGeNT (spectrum) &  $9.4/12$ & $\bar m = 1.0$ & $m_{F_2} = 36.0$ GeV, 
$\epsilon \sqrt{\xi_{F_2}} = 4.9 \times 10^{-9}$  \\
CRESST (spectrum) & $0.1/3$ & $\bar m = 1.0$ & $m_{F_2} = 80.0$ GeV, 
$\epsilon \sqrt{\xi_{F_2}} = 4.9 \times 10^{-9}$  \\
CDMS (spectrum) &   & $\bar m = 1.0$ & $m_{F_2} = 59.0$ GeV, 
$\epsilon \sqrt{\xi_{F_2}} = 2.4 \times 10^{-9}$  \\
\hline
DAMA (annual mod.) & $6.1/10$ & $\bar m = 3.0$ & $m_{F_2} = 58.0$ GeV, 
$\epsilon \sqrt{\xi_{F_2}} = 7.8 \times 10^{-9}$  \\
CoGeNT (spectrum) & $10.5/12$ & $\bar m = 3.0$ & $m_{F_2} = 44.0$ GeV, 
$\epsilon \sqrt{\xi_{F_2}} = 5.5 \times 10^{-9}$  \\
CRESST (spectrum) & $0.3/3$ & $\bar m = 3.0$ & $m_{F_2} = 74.0$ GeV, 
$\epsilon \sqrt{\xi_{F_2}} = 5.0 \times 10^{-9}$  \\
CDMS (spectrum) &   & $\bar m = 3.0$ & $m_{F_2} = 49.0$ GeV, 
$\epsilon \sqrt{\xi_{F_2}} = 2.4 \times 10^{-9}$  \\
\hline\hline
\end{tabular}
\caption{Summary of $\chi^2 (min)$ for the relevant data sets from the 
DAMA, CoGeNT, CRESST-II and CDMS/Si experiments for two reference $\bar m$ values and $v_{rot} = 200$ km/s. 
}
\end{table}

\begin{table}
\centering
\begin{tabular}{c c c c}
\hline\hline
Experiment & $\chi^2 (min)/d.o.f. $ & $\bar m$/GeV & Best fit parameters  \\
\hline
DAMA (annual mod.) & $5.0/10$ & $\bar m = 1.0$ & $m_{F_2} = 42.0$ GeV, 
$\epsilon \sqrt{\xi_{F_2}} = 7.4 \times 10^{-9}$  \\
CoGeNT (spectrum) &  $9.7/12$ & $\bar m = 1.0$ & $m_{F_2} = 28.5$ GeV, 
$\epsilon \sqrt{\xi_{F_2}} = 4.7 \times 10^{-9}$  \\
CRESST (spectrum) & $0.2/3$ & $\bar m = 1.0$ & $m_{F_2} = 57.7$ GeV, 
$\epsilon \sqrt{\xi_{F_2}} = 4.1 \times 10^{-9}$  \\
CDMS (spectrum) &   & $\bar m = 1.0$ & $m_{F_2} = 35.0$ GeV, 
$\epsilon \sqrt{\xi_{F_2}} = 2.1 \times 10^{-9}$  \\
\hline
DAMA (annual mod.) & $6.1/10$ & $\bar m = 3.0$ & $m_{F_2} = 46.0$ GeV, 
$\epsilon \sqrt{\xi_{F_2}} = 9.2 \times 10^{-9}$  \\
CoGeNT (spectrum) & $10.7/12$ & $\bar m = 3.0$ & $m_{F_2} = 36.5$ GeV, 
$\epsilon \sqrt{\xi_{F_2}} = 5.5 \times 10^{-9}$  \\
CRESST (spectrum) & $0.4/3$ & $\bar m = 3.0$ & $m_{F_2} = 39.8$ GeV, 
$\epsilon \sqrt{\xi_{F_2}} = 4.9 \times 10^{-9}$  \\
CDMS (spectrum) &   & $\bar m = 3.0$ & $m_{F_2} = 27.0$ GeV, 
$\epsilon \sqrt{\xi_{F_2}} = 2.3 \times 10^{-9}$  \\
\hline
\hline\hline
\end{tabular}
\caption{Summary of $\chi^2 (min)$ for the relevant data sets from the 
DAMA, CoGeNT, CRESST-II and CDMS/Si experiments for two reference $\bar m$ values and $v_{rot} = 240$ km/s. 
}
\end{table}

\vskip 0.5cm
\centerline{\epsfig{file=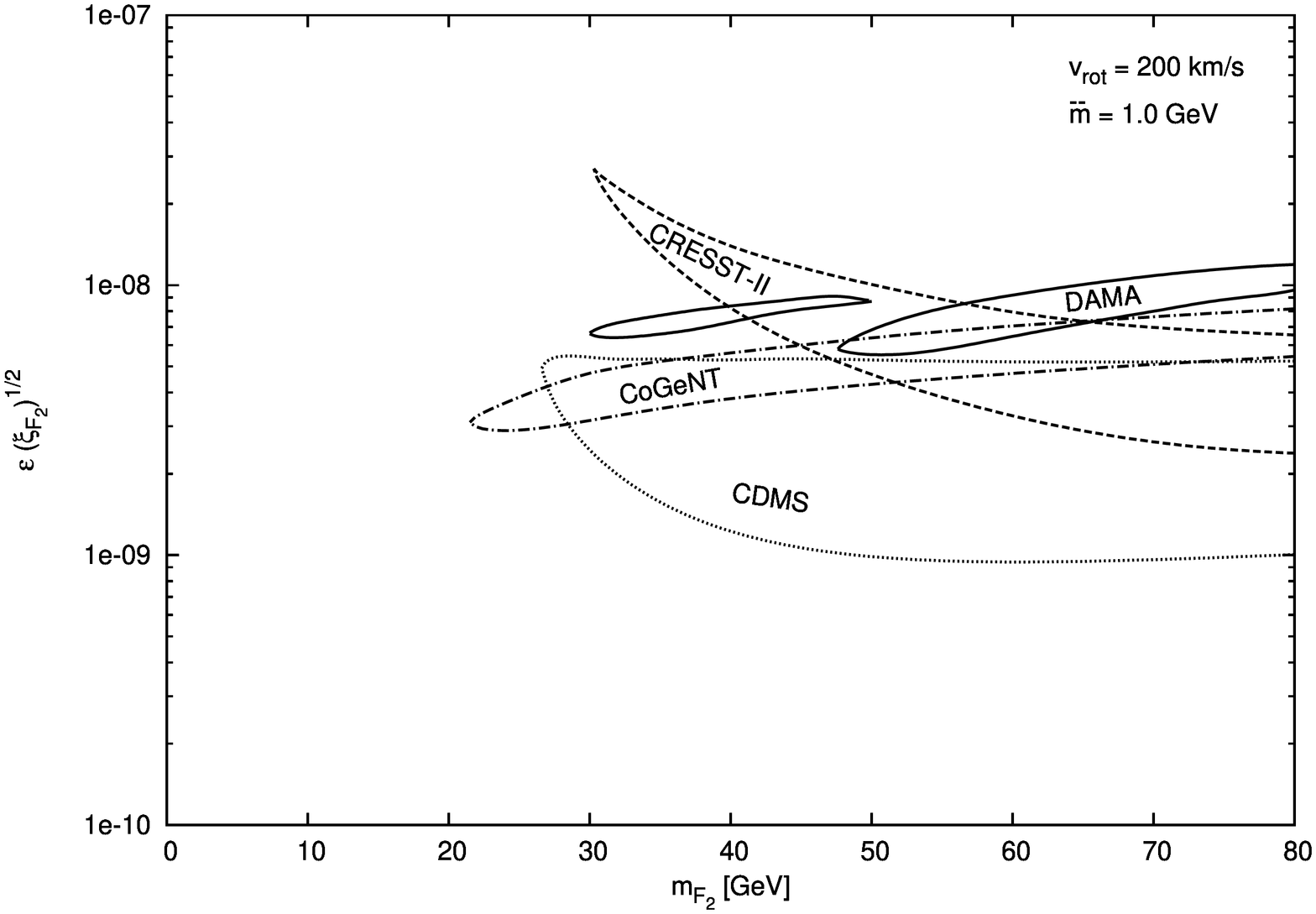,angle=270,width=13.0cm}}
\vskip 0.3cm
\noindent
{\small
Figure 1a: DAMA (solid line), CoGeNT (dashed-dotted), CRESST-II (dashed)  and CDMS/Si 
(dotted) favored regions of parameter space [all at 95\% C.L.] in the two component hidden
sector model.
The reference point $v_{rot} = 200$ km/s and $\bar m = 1.0$ GeV is assumed.
}

\vskip 0.2cm
\noindent

\vskip 0.1cm
\centerline{\epsfig{file=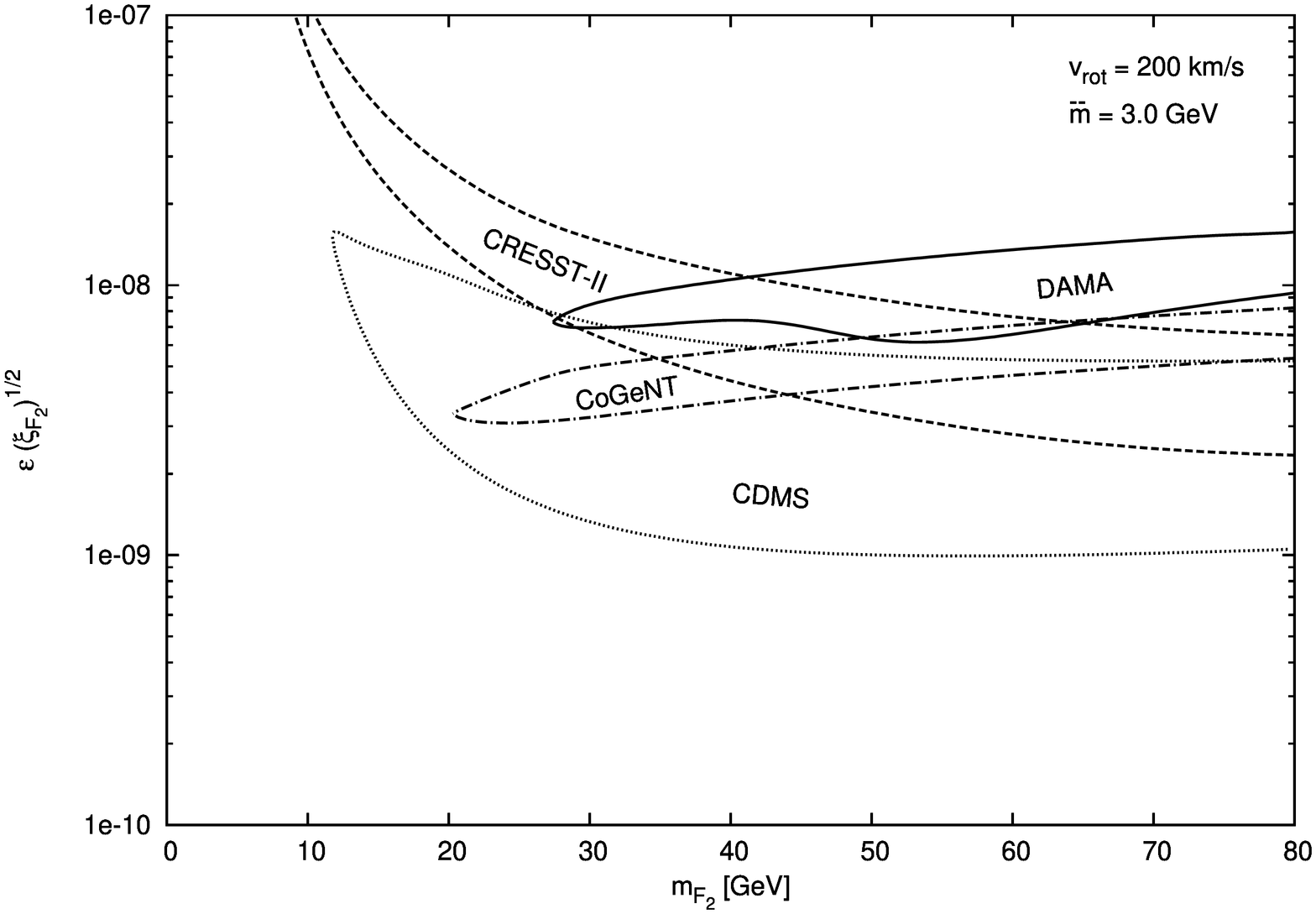,angle=270,width=13.0cm}}
\vskip 0.3cm
\noindent
{\small
Figure 1b: Same as figure 1a, except that $\bar m = 3.0$ GeV is assumed.
}

\vskip 0.4cm
\noindent
\vskip 0.5cm
\centerline{\epsfig{file=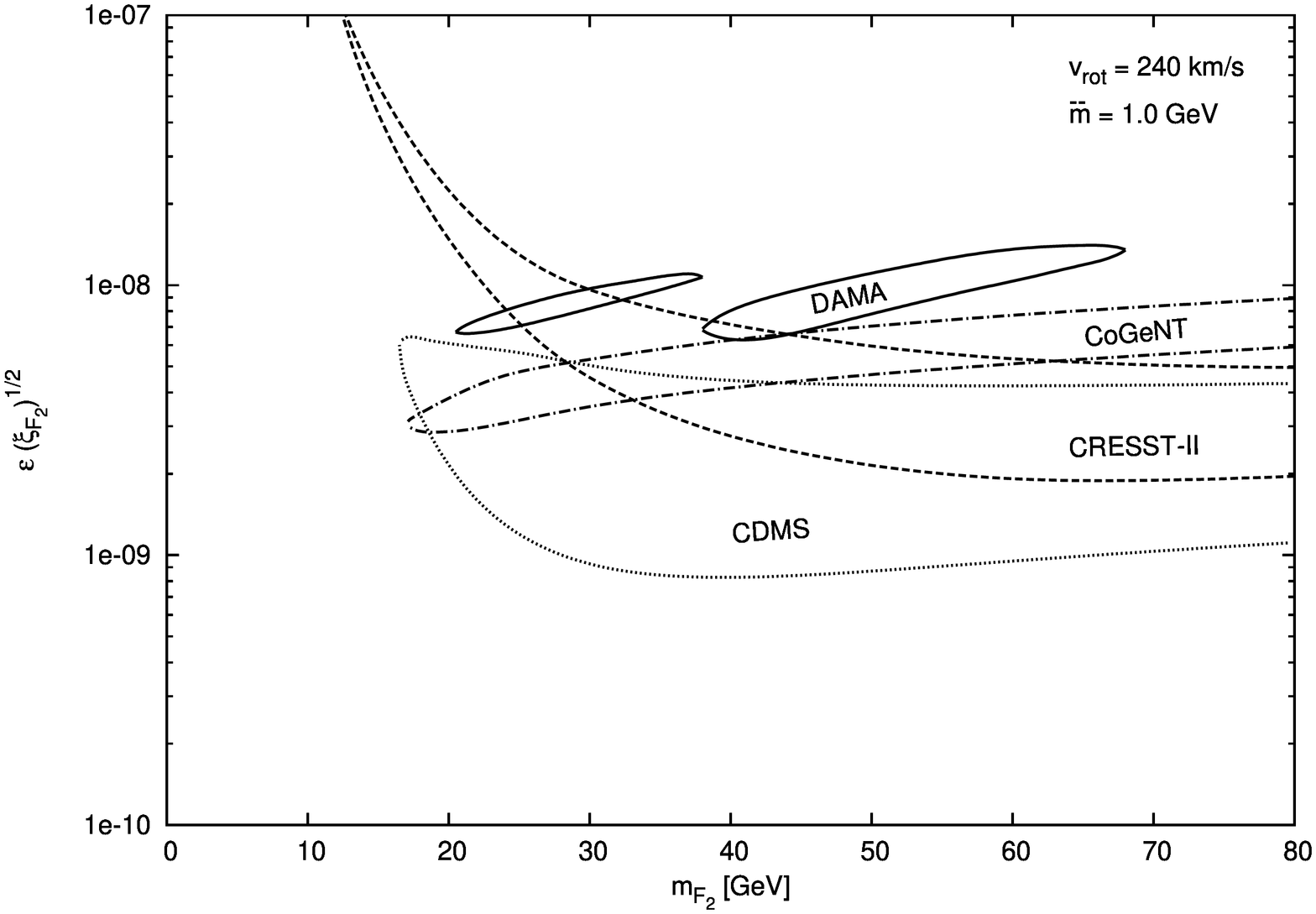,angle=270,width=13.0cm}}
\vskip 0.3cm
\noindent
{\small
Figure 2a: DAMA (solid line), CoGeNT (dashed-dotted), CRESST-II (dashed)  and CDMS/Si 
(dotted) favored regions of parameter space [all at 95\% C.L.] in the two component hidden
sector model.
The reference point $v_{rot} = 240$ km/s and $\bar m = 1.0$ GeV is assumed.
}

\vskip 0.5cm
\noindent

\vskip 0.5cm
\centerline{\epsfig{file=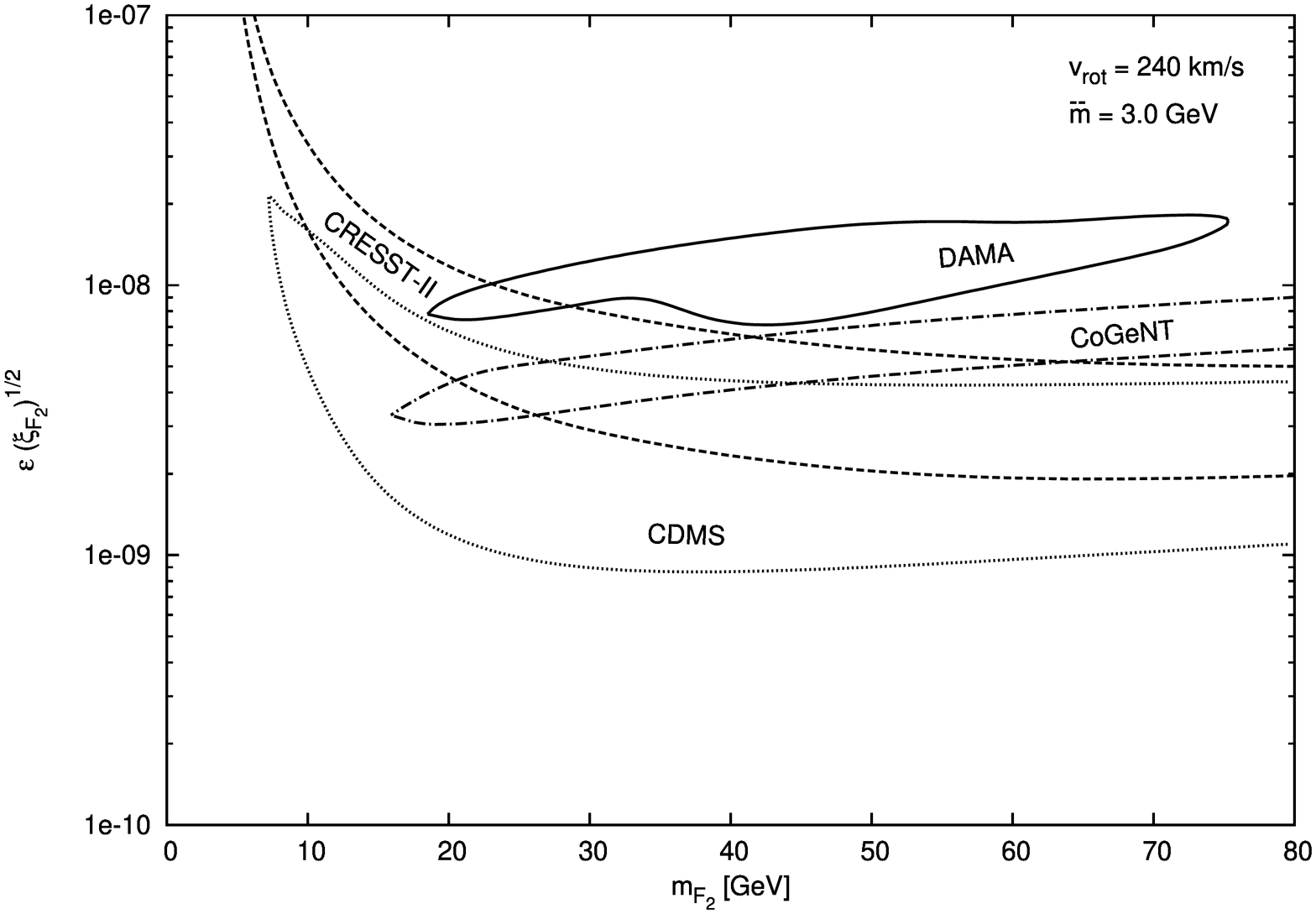,angle=270,width=13.0cm}}
\vskip 0.3cm
\noindent
{\small
Figure 2b: Same as figure 2a, except that $\bar m = 3.0$ GeV is assumed.
}

\vskip 0.8cm

The figures demonstrate that there are quite substantial regions of parameter space where each experiment can be
explained within this hidden sector framework. For DAMA, 
the signal is dominated by $F_2 - Na$ scattering if $m_{F_2} \stackrel{<}{\sim} 40$ GeV, 
while if $m_{F_2} \stackrel{>}{\sim} 40$ GeV then
both $F_2 - Na$ and $F_2 - I$ scattering conspire to produce the signal. 
Also, the signal arises from scattering
of target nuclei with
$F_2$ particles from the body of their 
Maxwellian halo distribution (rather than, say, the tail)
and thus is reasonably stable, and shows fairly mild dependence on the velocity dispersion i.e. on $\bar m$.

For CoGeNT, the spectrum is consistent with $dR/dE_R \propto 1/E_R^2$, predicted from the
energy dependence of the Rutherford cross-section. This is the reason why the CoGeNT spectrum is reproduced for 
a large range of $F_2$ mass. In other words, the shape of the CoGeNT spectrum arises from dynamics 
rather than kinematic effects
in this model. Again there is little dependence on the $\bar m$ parameter. 

In the case of CRESST-II, 
the dominant signal contribution arises from $F_2 - Ca$ scattering for
$m_{F_2} \sim 50$ GeV. For such masses,  
$F_2$ particles in the body of their Maxwellian halo distribution
can scatter to produce nuclear recoils above the 10 keV threshold.  
As the mass of $F_2$ is lowered, $m_{F_2} \stackrel{<}{\sim} 40-50$ GeV, 
a recoil above threshold can only occur for $F_2$ particles in 
the tail of the Maxwellian halo distribution.
For this reason, there is quite a bit of dependence of the CRESST-II allowed region on the velocity dispersion,
i.e. $\bar m$ in this mass range.   

The figures show that the allowed region of parameters favored by DAMA, CoGeNT, CRESST-II and CDMS/Si are very similar
with a significant
degree of overlap. This occurs ignoring a variety of possible systematic effects, such as the possibility
of channeled events in DAMA (which can lower the DAMA 
favored values of $\epsilon \sqrt{\xi_{F_2}}$), 
and uncertainties in the surface event correction factor for CoGeNT (which could raise or lower
the CoGeNT favored values of $\epsilon \sqrt{\xi_{F_2}}$).
Thus realistically there is quite a large region of parameter space that is possible, much larger than
the overlapping allowed region, even assuming all four experiments
have detected dark matter interactions. 
It is difficult, though,  to quantify all the possible systematic effects and do an exhaustive analysis.
Instead, we shall hope to gain some insight by examining a particular point in parameter space. 

\section{An example near the combined best fit of DAMA, CoGeNT, CRESST-II and CDMS/Si}

In view of potential systematic uncertainties between the experiments we
shall resist the temptation to fit the combined DAMA, CoGeNT, CRESST-II and CDMS/Si data.
Instead we shall consider an
example reference point $P1$ located near the overlapping allowed regions
indicated in figure 1a.
\begin{eqnarray}
P1: m_{F_2} &=& 50 \ {\rm GeV}, \ \epsilon \sqrt{\xi_{F_2}} = 5.7\times
10^{-9}, \ \bar m = 1.0\ {\rm GeV}, \ v_{rot} = 200\ {\rm km/s}
\ .
\nonumber 
\end{eqnarray} 
The DAMA annual modulation signal for this example is given in figure 3a,
the CoGeNT spectrum in figure 3b, and CRESST-II spectrum in figure 3c.

\vskip 0.5cm
\centerline{\epsfig{file=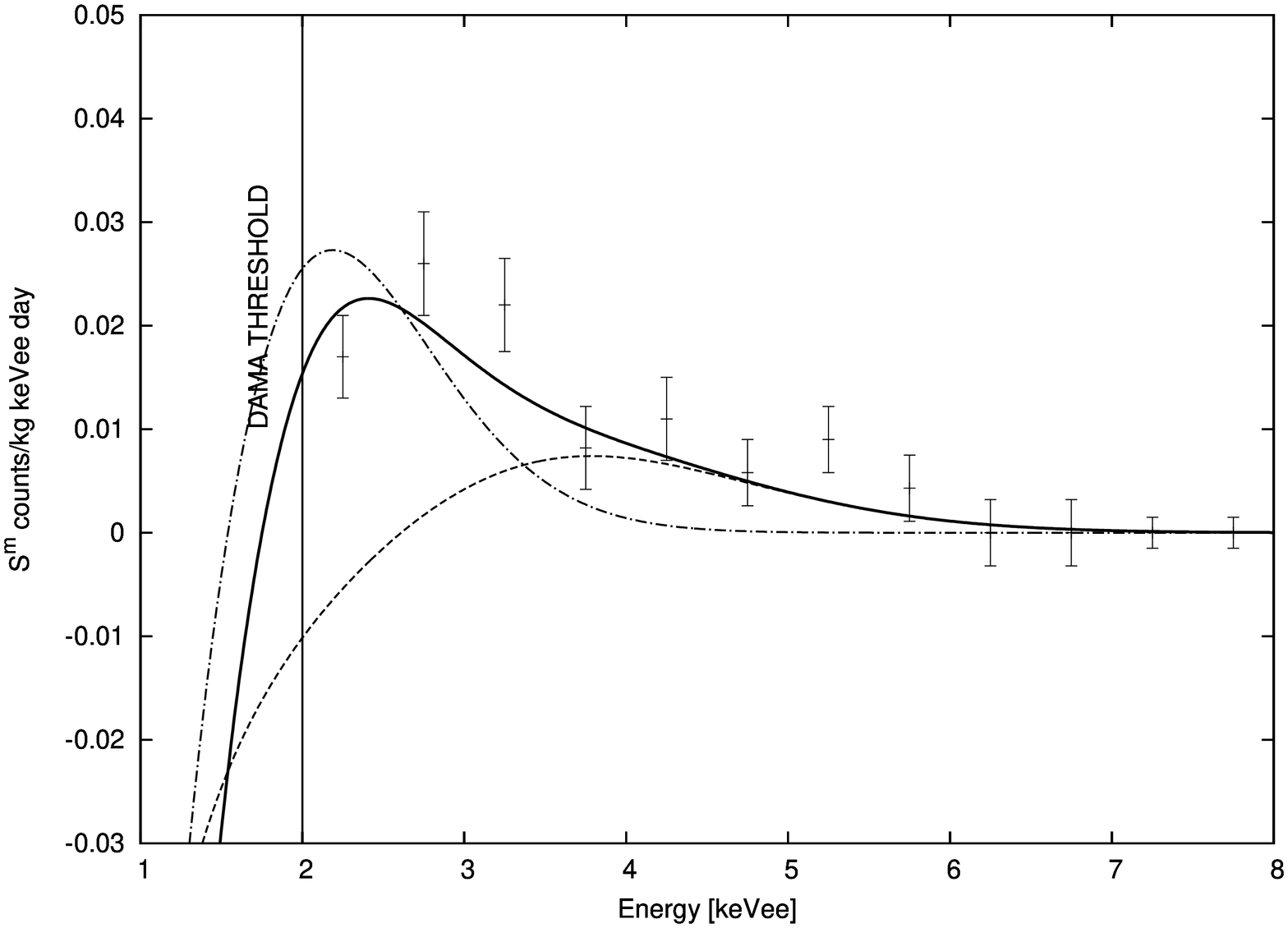,angle=270,width=12.7cm}}
\vskip 0.3cm
\noindent
{\small
Figure 3a: DAMA annual modulation spectrum for two component hidden sector dark matter with 
parameters $P1$ (solid line).  
The separate contributions from scattering off sodium (dotted line) and Iodine (dashed-dotted line)
are also shown.  
In this example $q_{Na} = 0.36, \ q_I = 0.20$.}

\vskip 0.5cm
\centerline{\epsfig{file=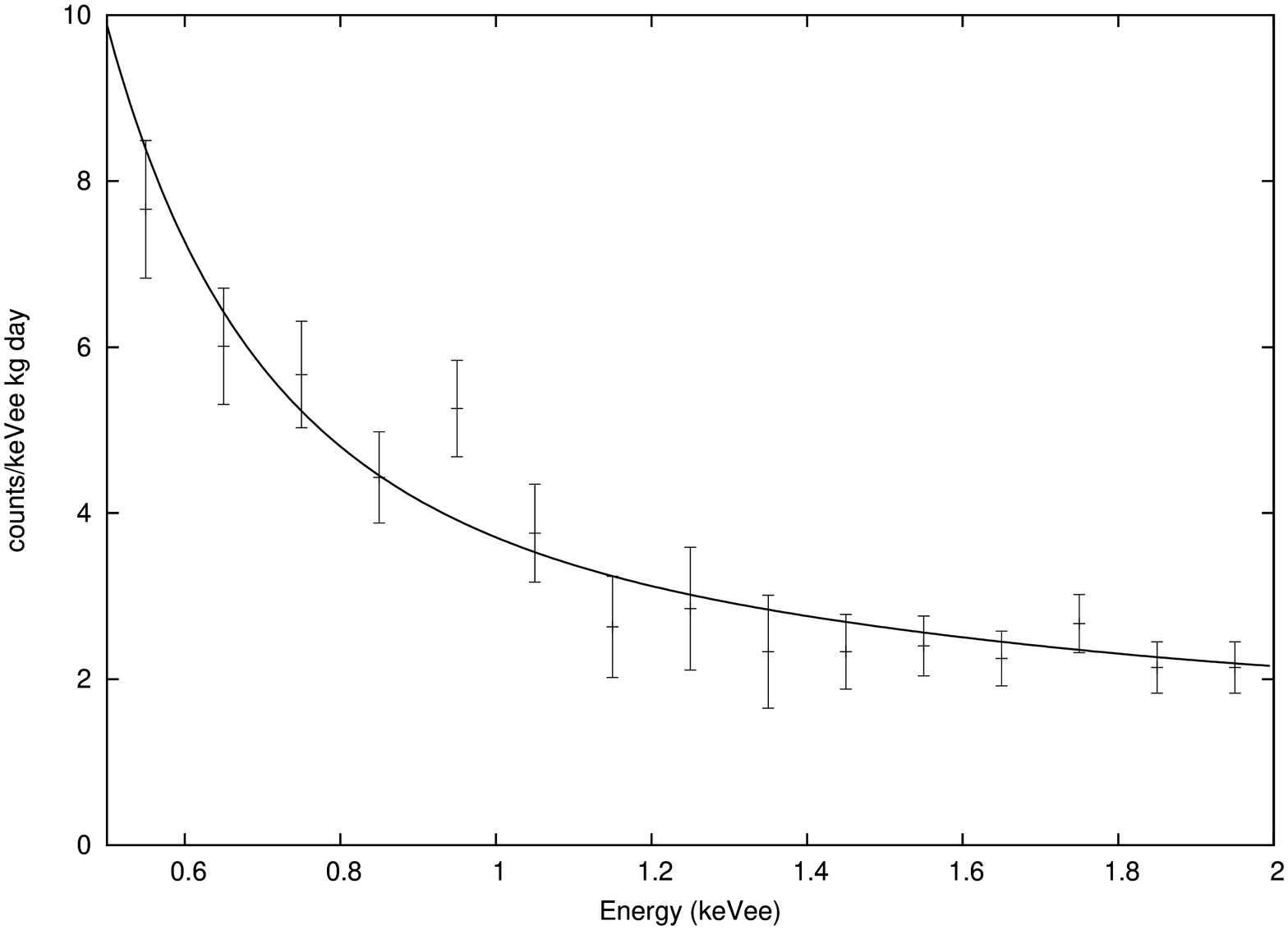,angle=270,width=12.7cm}}
\vskip 0.3cm
\noindent
{\small
Figure 3b: CoGeNT spectrum for two component hidden sector dark matter with 
parameters $P1$.
In this example $q_{Ge} = 0.17$.}

\vskip 0.5cm
\centerline{\epsfig{file=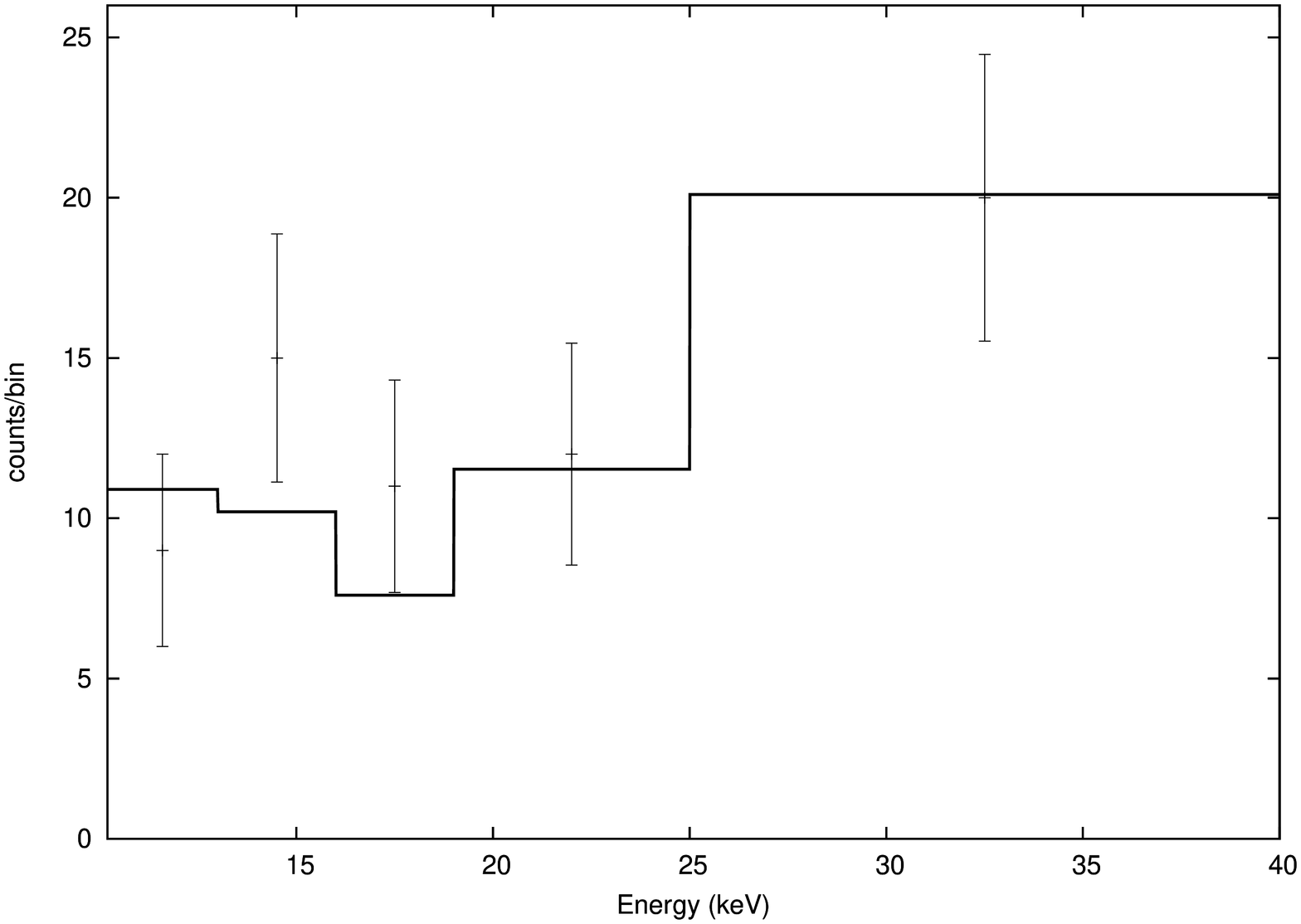,angle=270,width=13.0cm}}
\vskip 0.3cm
\noindent
{\small
Figure 3c: CRESST spectrum for two component hidden sector dark matter for the example $P1$. 
}
\vskip 1.2cm

Figures 3 clearly demonstrates that two component hidden sector dark matter can 
simultaneously explain DAMA, CoGeNT and CRESST-II data.
The apparent bump in the DAMA data can be reproduced extremely well and the minimal model
predicts a change in sign of the DAMA annual modulation signal at low energies. Note though,
that if there are additional particles,  
$F_3$, of intermediate mass $\sim 20$ GeV, then their positive contribution
to the annual modulation can outweigh the negative contribution from $F_2$.
An example, given in the context of mirror dark matter with $F_2 \sim Fe'$ and $F_3 \sim O'$,
was discussed in ref.\cite{foot2012}.
Figure 3b indicates that
the shape of the CoGeNT spectrum is consistent with the $dR/dE_R \propto 1/E_R^2$
energy dependence of the Rutherford cross-section.
The event rate is predicted to continue to rise as $\sim 1/E_R^2$ as the recoil energy is reduced below the
current
CoGeNT threshold, until the threshold of a lighter component is reached, whereby the rate can jump even higher.
These effects can be probed by TEXONO, C-4, CDMSlite and possibly other experiments. 

Future data from DAMA, CoGeNT, CRESST-II and other experiments will obviously be able
to constrain the parameter space within this hidden sector 
framework.  As discussed recently\cite{diurnal},
a particularly striking diurnal modulation signal should be observable
for a detector located in the southern hemisphere, and perhaps even in
the northern hemisphere at low
latitudes, such as detectors in Jin-Ping underground laboratory.
In the meantime, we must rely on annual modulation and spectrum data.

In figure 4 we give the predicted spectrum
for DAMA/LIBRA for the reference point, $P1$.
Figure 4 also shows the `single hit'
event rate recorded in DAMA/LIBRA\cite{dama2}.
The figure indicates that the sharp rise in the predicted dark matter interaction rate  
could potentially be differentiated from the background  
if the DAMA threshold is lowered below $2$ keVee.
In figure 5
we show the predicted
annual modulation spectrum for CoGeNT for the same reference point,
$P1$. 
Clearly, the initial annual modulation amplitude
measured by CoGeNT to be $A \approx 0.46 \pm 0.17$ cpd/kg/keVee averaged over the energy
range: $0.5 < E({\rm keVee}) < 3.0$, is much larger than
that predicted by our example point. 
Also, we find that the annual modulation changes sign at low energies. This feature
is not supported by CoGeNT's initial measurement.
However the energy where the modulation changes sign can be reduced if the mass of $F_2$ is 
lowered, which
can alleviate this discrepancy.
Alternatively, there can be an additional dark component, $F_3$, of mass $\sim 20$ GeV. As discussed earlier in
the context of DAMA, 
such a component can give a positive contribution
to the annual modulation which can outweigh the negative contribution from $F_2$.
Clearly future measurements of the annual modulation by CoGeNT,
C-4, CDEX, and other experiments will be very important.

\vskip 0.5cm
\centerline{\epsfig{file=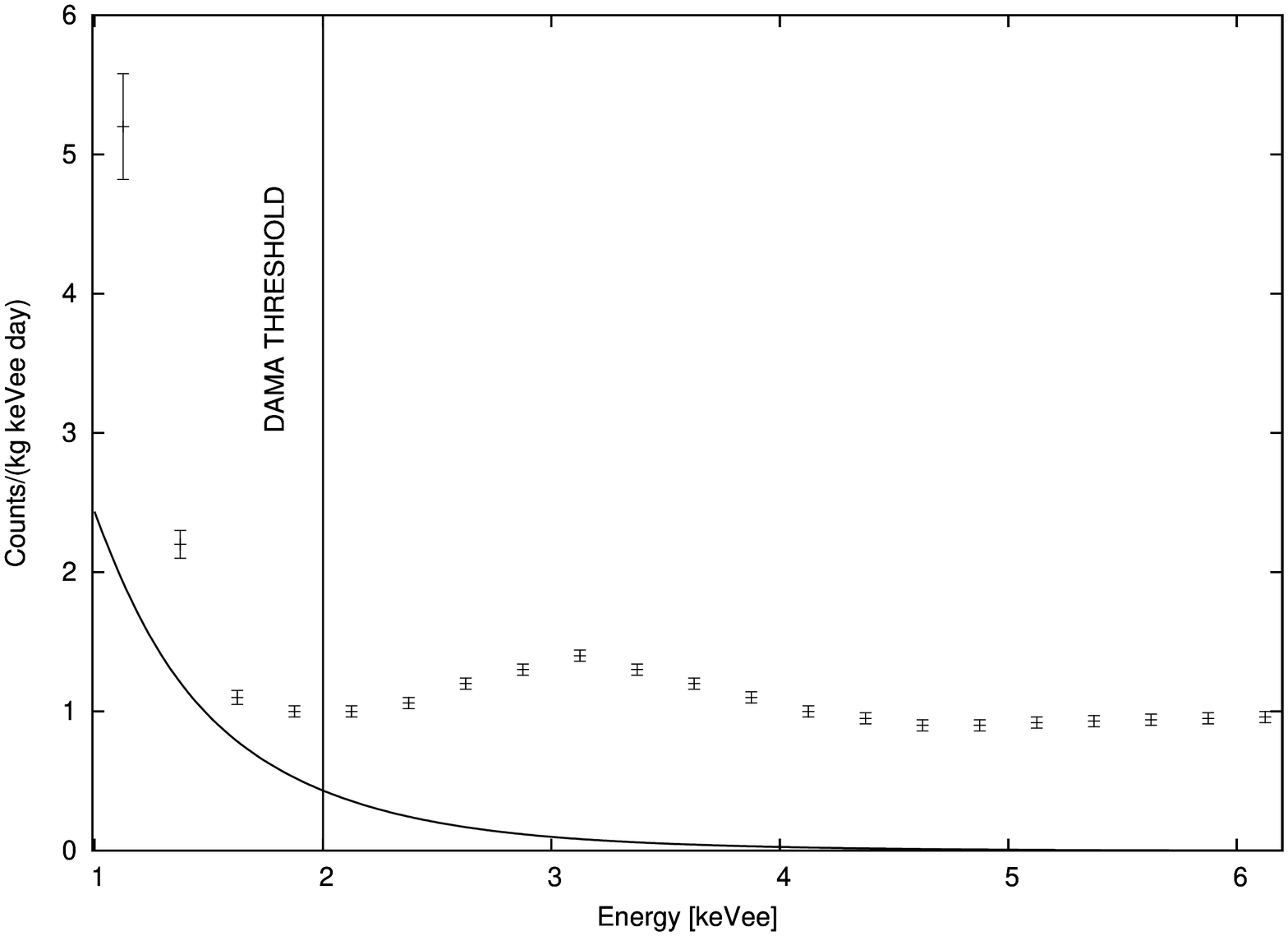,angle=270,width=13.0cm}}
\vskip 0.3cm
\noindent
{\small
Figure 4:
DAMA spectrum for hidden sector dark matter with
parameters $P1$.
In this example $q_{Na} = 0.36$, $q_I = 0.20$.}
\vskip 0.7cm

\vskip 0.5cm
\centerline{\epsfig{file=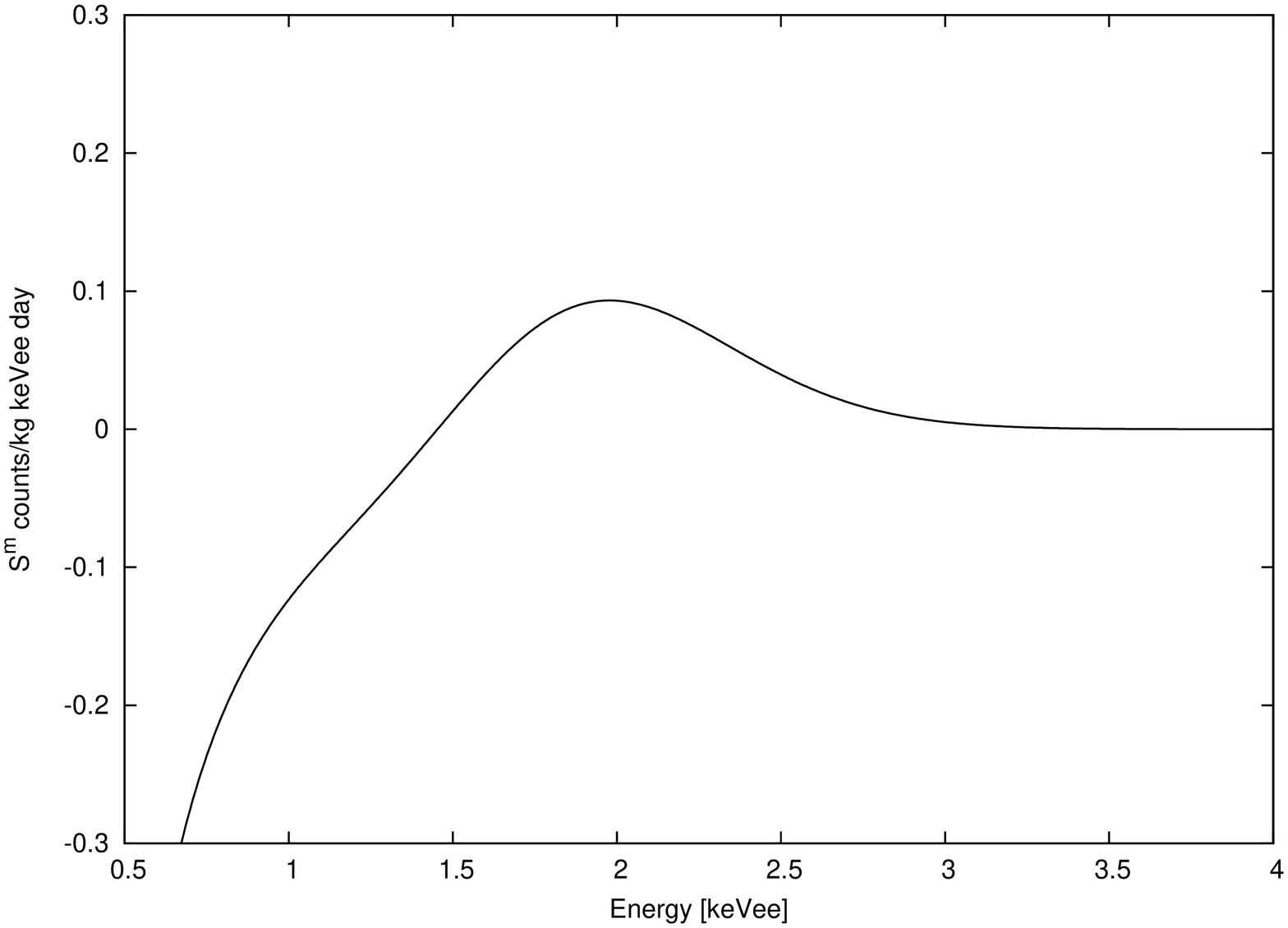,angle=270,width=13.0cm}}
\vskip 0.3cm
\noindent
{\small
Figure 5:
CoGeNT annual modulation spectrum for hidden sector dark matter with
parameters $P1$.  In this example $q_{Ge} = 0.17$.
}
\vskip 0.8cm

\section{XENON100 constraints}

The XENON100\cite{xenon100}
experiment has reported null results in their dark matter search.
Here, we examine the 
compatibility of the considered hidden sector dark matter model with
these null results
\footnote{
There are also lower threshold analysis by the XENON10\cite{xenon10} and
CDMS collaborations\cite{cdms}.
However it has been argued\cite{collarguts} that neither analysis can
exclude light dark matter (and by extension, hidden sector dark matter examined here, which
has similar event rates at low energies)
when systematic uncertainties are properly
taken into account. Interestingly, a recent analysis\cite{collarf} has
found that the low energy CDMS data are actually fully consistent with
CoGeNT's observed low energy excess rate, adding weight to the dark matter interpretation 
of this excess.}.
The constraints from the XENON100 experiments
depend sensitively on the
recoil energy threshold, which has significant uncertainty. The precise
magnitude of this uncertainty is unknown, but the subject of active discussions\cite{collarzzz,damaguts}.
Given this situation, we examine the compatibility issue by 
estimating the energy threshold
for which the parameter point $P1$ can be excluded at $95\%$ C.L. 
Taking into account the relevant
detection efficiencies, exposure time and detector resolution, we find that
the energy threshold of the XENON100 experiment needs to be around 15 keV for the point
$P1$ to be consistent at $95\%$ C.L. This can be compared with the nominal threshold energy
of 6.4 keV. 
Our example point assumed $\bar m = 1.0$ GeV. Raising $\bar m$ will increase the
tension of hidden sector dark matter with the null results the XENON100 experiment.
Lowering $\bar m$ can improve the situation, but only moderately so. 
Although $P1$ was just an example point, it seems that some level
of tension exists between the null results of XENON100 and the hidden sector dark matter
explanation of the DAMA, CoGeNT, CRESST-II and CDMS/Si experiments. 

One can envisage several possible ways in which this tension might be alleviated.
For example, it is possible that there is an issue with
the calibration of the XENON100 apparatus. Determining the recoil energy scale in the XENON100 
detector is nontrivial and it seems possible that this scale might be 
have a factor $\sim 2$ uncertainty\cite{collarzzz,damaguts}.
Another possibility is that $F_2$ has a somewhat lower
mass than given in our example. 
This option would be especially relevant if we were to ignore the CRESST-II excess\footnote{
See \cite{kuz} for a discussion of a subtle background effect which might potentially
explain the CRESST-II low energy excess.}.
For example
the parameter point:
\begin{eqnarray}
P2: m_{F_2} &=& 40 \ {\rm GeV}, \ 
\epsilon \sqrt{\xi_{F_2}} = 5.7\times 10^{-9}, \ \bar m = 1.0\ {\rm GeV}, \ v_{rot} = 190\ {\rm km/s}
\nonumber 
\end{eqnarray} 
yields a $\chi^2 (dama) = 18.9$ for 12 data points and $\chi^2(cogent) = 22.8$ for 15 data points, but gives
only around 1 event for the CRESST-II exposure. 
This is a reasonable fit for DAMA and CoGeNT, considering that the shape of the distributions are fit well,
only the overall normalization is not [CoGeNT (DAMA) prefers slightly smaller (larger) 
$\epsilon\sqrt{\xi_{F_2}}$, with $\epsilon \sqrt{\xi_{F_2}} = 5.7 \times 10^{-9}$ a compromise]. 
The different normalizations might easily be due to
systematic effects not included in our analysis.
Taking into account the relevant
detection efficiencies, exposure time and detector resolution, we find that
the energy threshold of the XENON100 experiment needs to be around 11.5 keV for the point
$P2$ to be consistent at $95\%$ C.L.\footnote{
We have checked that the example points $P1$ and $P2$ are consistent with 
the CDMS/Ge\cite{cdmsge} data (taking a systematic uncertainty in energy scale of 20\%)
and also the KIMS experiment\cite{kims}.
}.

\section{Conclusion}

We have examined the data from the DAMA, CoGeNT, CRESST-II and CDMS/Si experiments 
in the context of multi-component hidden sector dark matter. The models considered
feature a hidden sector with two or more stable particles charged under an unbroken $U(1)'$ gauge interaction.  
The new gauge field can interact with
the standard $U(1)_Y$ via renormalizable kinetic mixing, leading to Rutherford-type elastic
scattering of the dark matter particles off ordinary nuclei.
We examined the simplest generic model of this type, with a hidden sector composed of two stable 
particles, $F_1$ and $F_2$. 

We have found that the two component hidden sector dark matter model can simultaneously explain the
DAMA, CoGeNT, CRESST-II and CDMS/Si data.
This explanation has some
tension with the XENON100 experiment. The favored parameter regions
are typically consistent with the most recent XENON100 results only if the
XENON100 energy threshold is around a factor of two higher than given by
the XENON100 collaboration.

\vskip 2cm
\noindent
{\large \bf Acknowledgments}

\vskip 0.2cm
\noindent
This work was supported by the Australian Research Council.

\end{document}